# MINIMUM PROCESS COORDINATED CHECKPOINTING SCHEME FOR AD HOC NETWORKS

RuchiTuli[1]&Parveen Kumar[2]

[1]Research Scholar, Singhania University (Pacheri Bari), Rajasthan (INDIA)
Present Address : Lecturer, Yanbu University College, Yanbu (Saudi Arabia)
`tuli.ruchi@gmail.com`

[2]Professor, Meerut Institute of Engineering & Technology, Meerut (INDIA)
`pk223475@yahoo.com`

**ABSTRACT**

*The wireless mobile ad hoc network (MANET) architecture is one consisting of a set of mobile hosts capable of communicating with each other without the assistance of base stations. This has made possible creating a mobile distributed computing environment and has also brought several new challenges in distributed protocol design. In this paper, we study a very fundamental problem, the fault tolerance problem, in a MANET environment and propose a minimum process coordinated checkpointing scheme. Since potential problems of this new environment are insufficient power and limited storage capacity, the proposed scheme tries to reduce the amount of information saved for recovery. The MANET structure used in our algorithm is hierarchical based. The scheme is based for Cluster Based Routing Protocol (CBRP) which belongs to a class of Hierarchical Reactive routing protocols. The protocol proposed by us is non-blocking coordinated checkpointing algorithm suitable for ad hoc environments. It produces a consistent set of checkpoints; the algorithm makes sure that only minimum number of nodes in the cluster are required to take checkpoints; it uses very few control messages. Performance analysis shows that our algorithm outperforms the existing related works and is a novel idea in the field. Firstly, we describe an organization of the cluster. Then we propose a minimum process coordinated checkpointing scheme for cluster based ad hoc routing protocols.*

**KEYWORDS**

*Ad hoc routing, checkpointing, fault tolerance, mobile computing, clusterheads, clustering routing protocol.*

## 1. INTRODUCTION

With recent advances in mobile technology and mobile devices, mobile computinghas become an important part of our life. People are using wireless networks fortheir day-to-day work, be it making a phone call or to download news or to see andlisten or only listen to their favorite song from various multimedia servers with thehelp of various devices such as mobile phones, PDAs or a laptop. More servicesare in the offering in near future. The desire to be connected anytime,anywhere, anyhow has led to the development of wireless networks, opening newvista of research in pervasive and ubiquitous computing. This emerging fieldof mobile and nomadic computing requires a highly failure free environment to effectively manage the communication among the peers.

Ad hoc networks have recently been considered as an attractive research filed. In some case, such as emergency, disaster relief or battlefield operations, when a wire line is not available, an ad hoc network can be set for the communication. Clustering of MH provides a convenient framework for resource management. The main advantage of clustering is reducing the number of messages sent to each BS from each node, channel access, power control and bandwidth control. In cluster based architecture, whole network is divided into several clusters and in each cluster network elects one node to be called as cluster head. Hence,clustered ad hoc network consists of three kinds of nodes – cluster heads, gateways and ordinary nodes. Clusterheads are the nodes that are given the responsibility for routing the messages within the cluster and performing the data aggregation. The communication between two adjacent clusters are conducted through the gateway nodes. All nodes other than that gateway and clusterheads are called ordinary nodes. Both gateways and ordinary nodes are managed by their clusterheads.

There is no physical backbone architecture available in ad hoc wireless networks. for routing of the message, a node depends on other nodes to relay packets if they do not have direct links. Wireless backbone architecture can be used to support efficient communications between nodes [1], [2], [3], [5]. To support backbone architecture, the clusterheads should be a part of the backbone and the fewer the number of backbone nodes the better. Fewer nodes in the backbone can reduce the quality of messages exchanged by backbone nodes [3], [4]
.
In this paper, we propose a checkpointing scheme for clustering routing protocol as a method of improving reliability. A cluster head send routing and collected data information to BS, which periodically save the state of cluster head. If a cluster head fails or some fault is detected, then BS detects the cluster head failure and some new node in the cluster is assigned the responsibility of the cluster head. Using checkpointing the cluster can quickly recover from a transient fault of cluster head. The merits of our work are as follows. .We propose a minimum process checkpointing algorithm for cluster based architectures in which a MH first takes a tentative checkpoint and later on when it receives commit request from the initiator, MH converts its tentative checkpoint into permanent checkpoint.The paper is organized as follows. Section 2 discusses the background material for this work. We give System Design in Section 3 and Section 4 provides the organization and setup of a cluster. The proposed checkpointing scheme is formulated in Section 5. Section 6 discusses handling of disconnections and Section 7 shows the performance of our algorithm. Finally, Section 8 concludes the paper.

## 2. Related Work and problem formulation

### 2.1 Related work

In this section we briefly introduce prior studies related to our work.

In [6], the authors proposed the concept of in-network fault tolerance for achieving enhanced network dependability and performance. In that scheme, the sink node periodically checkpoints its state and saves it in the memory of one or more sensor nodes, so called checkpoint sensors. When a sink node ($S_1$) fails or reaches an energy level below its threshold, another sensor node will be selected to operate as the new sink node ($S_2$). After applying this approach m times, the sink will be located in a sensor denoted by Sm. If the sink is located on $S_m$, then $S_{m-1}$ is the checkpoint sensor and the path between S1 and $S_m$ is the checkpoint path. When a sink node ($S_m$) fails, $S_{m-1}$ detects the failure and becomes the sink instead; it iteratively operates in this sequence through the checkpoint path. This scheme is simple to implement, but energy consumption and reliability vary according to the position of the sink node.

In [7], a cluster takes two types of checkpoints – processes inside the cluster take synchronous checkpoints and a cluster takes a communication induced checkpoint whenever it receives an inter-cluster application message. Each cluster mainitains a sequence number (SN). SN is incremented each time a cluster level message is committed.

In [8], authors proposed a simple non-blocking roll-forward checkpointing/recovery mechanism for cluster federation. The main feature of their algorithm is that a processs receiving a message does not need to worry whether the received message may become orphan or not. It is the responsibility of the sender of the message to make it non-orphan.

In [22], the authors proposed a integrated independent and coordinated checkpointing schemes for the applications running in hybrid distributed environments. They stated that independent checkpoint subsystem takes a new coordinated checkpoint set if it sends an intercluster application message. Also a process pi of independent checkpointing subsystem takes a new independent checkpoint before processing an already received intercluster application message, if pi has sent any intracluster application message after taking its last checkpoint.

In [9] and [10] the authors have proposed non-blocking coordinated checkpointing algorithms that require minimum number of processes to take checkpoints at any instant of time.

A good checkpointing protocol for mobile distributed systems should have low overheads on MHs and wireless channels and should avoid awakening of MHs in doze mode operation. The disconnection of MHs should not lead to infinite wait state. The algorithm should be non-intrusive and should force minimum number of processes to take their local checkpoints [11]. In minimum-process coordinated checkpointing algorithms, some blocking of the processes takes place [12], [13], or some useless checkpoints are taken [9], [10], [14].

Cao and Singhal [9] achieved non-intrusiveness in the minimum-process algorithm by introducing the concept of mutable checkpoints. The number of useless checkpoints in [9] may be exceedingly high in some situations [14]. Kumar et. al [14] and Kumar et. al [10] reduced the height of the checkpointing tree and the number of useless checkpoints by keeping non-intrusiveness intact, at the extra cost of maintaining and collecting dependency vectors, computing the minimum set and broadcasting the same on the static network along with the checkpoint request.

Higaki and Takizawa [15] proposed a hybrid checkpointing protocol where the mobile stations take checkpoints asynchronously and fixed ones synchronously. Kumar and Kumar [17] proposed a minimum-process coordinated checkpointing algorithm where the number of useless checkpoints and blocking are reduced by using a probabilistic approach. A process takes its mutable checkpoint only if the probability that it will get the checkpoint request in the current initiation is high. To balance the checkpointing overhead and the loss of computation on recovery, P Kumar [16] proposed a hybrid-coordinated checkpointing protocol for mobile distributed systems, where an all-process checkpoint is taken after executing minimum-process checkpointing algorithm for a certain number of times.

In this paper, we have proposed a non-blocking minimum process checkpointing scheme for ad hoc networks which makes sure that only minimum number of nodes in a cluster are required to take checkpoint in the execution of checkpointing algorithm . We have developed our scheme for a class of Cluster Based Routing Protocols (CBRP) in ad hoc networks. The schemes described above consider a general model of mobile networks.

## 2.2 Problem Formulation

The mobile ad hoc network distinguishes itself from traditional wireless networks by its dynamic changingtopology, no base station support and the need of multihop communication MANET, a mobile host (MH) isfree to move around and may communicate with others at anytime. When a communication partner is within a host's radio coverage, they can communicate directly with each other in a one-hop manner. Otherwise, aroute consisting of several relaying hosts is needed to forward messages from the source to the destination ina multihop fashion. However, in order to construct the routing path, the source host needs to send the request to all its neighbors. On receiving the request message, the neighboring host has to check whether it is the destination or not. If not, it should continue relaying the request packet to all it neighbors until the request packet reaches the destination. In this situation, it will cause packet flooding. Thus, in order to reduce the flooding packets and minimize the data of routing table, the cluster_based (or hierarchically organized) model is proposed to alleviate the flooding phenomenon. Clustering an ad hocnetwork means partitioning its nodes into clusters CLs, each one with a clusterhead (CH) and possibly some ordinary nodes. The clusterhead which acts as a local coordinator of transmissions within the cluster.Each cluster is represented by the ID of its clusterhead. For example, Figure 1 shows a cluster baseddistributed mobile computing systems and there are four clusters CL1, CL2, CL3 and CL4. A MH can communicate with other MHs in different cluster or in the same cluster only through its own CH. A clustered architecture is characterized by two types of messages – inter-cluster messages and intra-cluster message.The main aim of clustering routing protocols is to efficiently maintain energy consumption of nodes by involving them in multi-hop communication within a particular cluster and by performing data aggregation in order to decrease the number of messages transmitted to MSS. Since Normal nodes only communicate with their cluster head, which in turn, aggregates the collected information and sends it to the MSS. In this scheme, cluster head failures are more critical than those normal nodes. When a cluster head fails, re-election of cluster head is performed within the cluster. Such a recovery scheme is a time and energy consuming process. Therefore, to improve the quality and reliability of ad hoc networks, a fault tolerant mechanism is needed for such cluster heads. In this paper, we propose a checkpointing scheme for the cluster based ad hoc networks.

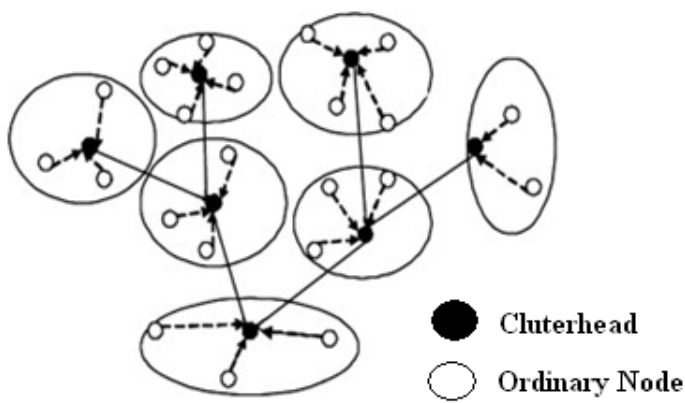

**Figure 1 : The concept of a clustering routing protocol**

During the cluster head election setup, our scheme elects the cluster head that has more weight function [23]. Then we have proposed a non-blocking coordinated checkpointing algorithm in which MHs take a tentative checkpoint and then on receiving a commit message from the initiator, the MHs convert their tentative checkpoint into permanent. Also whenever a MH is busy, the process takes a checkpoint after completing the current procedure. The proposed algorithm requires fewer control messages and hence fewer number of interrupts. Also, our algorithm requires only minimum number of MHs in a cluster to take checkpoints, it makes our algorithm suitable for cluster based protocols in ad hoc networks.

## 3. System design

The protocols of mobile distributed environment are not suitable for ad hoc environments due to their different architectures. For an algorithm to work feasible in an ad hoc environment, it should satisfy the following properties :-
   a) Every ordinary node must have at least one cluster head as a neighbour (dominance property)
   b) No two clusterheads can be neighbours (independence property)
   c) In a cluster, any two nodes are at most two hops away, since the clusterhead is directly linked to every node in a cluster (two-hop property)

Based on these properties each node is either a clusterhead or is directly linked to one or more clusterheads and each clusterhead can take and maintain control of its members efficiently[13].

### 3.1 Notations and assumptions

Most hierarchical clustering architectures are based on cluster head concept. The cluster head acts as a local coordinator within each cluster and it resembles a base station in cellular systems. The details of the notations and assumptions are described as follows. According to fig. 2, the system topology can be expressed as a graph G=(V, E), where V is the set of nodes and E is the set of edges. The symbols and definitions that we have used are as follows :

**n:** it denotes the total number of nodes in network. In other words, n=|V|.
**status :** it denotes the nodes role. There are two types of status CH and O where CH denotes that node is a cluster head and O denotes that the nodes is an ordinary node or gateway.

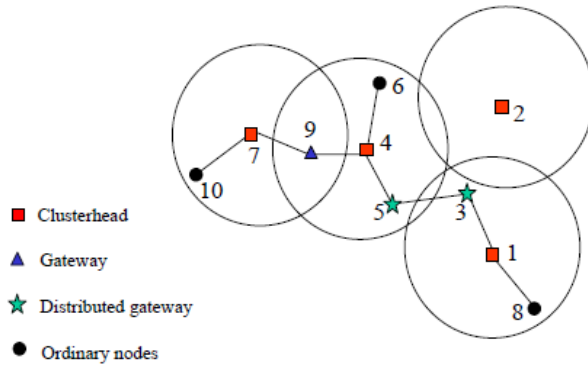

**Figure 2 : A cluster-based wireless network**

### 3.2 Electing a cluster head

There are five well known clustering algorithms to elect a cluster head. The first scheme is highest connectivity [13], in which the node with the highest degree is always selected as the cluster head by the adjacent nodes in the same cluster. The major disadvantage of this algorihm is the frequent cluster head change problem.

International Journal on AdHoc Networking Systems (IJANS) Vol. 1, No. 2, October 2011

The second scheme is the lowest-id clustering algorithm [18], in which each node is assigned a unique id. The node with the lowest id is always selected as the cluster head.

The third scheme is the least cluster change clustering (LCC) algorithm [19], which restricts cluster head changes under two conditions. The first condition is when two cluster heads come within transmission range of one another and the second one when a node becomes disconnected from any other cluster.

The fourth one is distributed mobility adaptive clustering (DMAC) [20], in which nodes are grouped using a weight-based criterion. The choice of clusters is based on a generic weight associated with each node : the larger the weight, the better the node fits the role of a cluster head.

The final algorithm is weighted highest degree clustering algorithm [21], in which a new weight function is used to select the cluster heads. The weight function is defined as the sum of the reciprocal of the neigbours degree

We use the weight function in our scheme to elect the cluster head. We define the weight function here.

*Weight(x)* is a function that returns the weight of the nodex. We use weight function here to determine if the node is an inner node or a border node. In general the inner nodes have higher weight values that the outer nodes. The weight of a node is defined as :
*Weight (x) = $\Sigma$Degree (y) + ID(x)/n+1, y $\in$ N(x)*
*Where,*
***Degree(x)****is a function that returns the degree of a node. For instance, in figure 2, Degree of nodes 1, 2, 3 and 4 are 2, 0, 2 and 3 and
**ID(x)** is a function the returns the id of a node x. Each node in ad hoc network has a unique id. For instance, the ID number of nodes 1, 2, 3, and 4 are 1, 2, 3 and 4.

Since each node has a unique ID number, so no two nodes can have the same weight function even though the degree of two nodes can be same.

### 3.3 System model

The system model consists of a network which is divided into clusters and each cluster has 1 cluster head**,** rest all the nodes are ordinary nodes. The following assumptions are made in our system model :
   a) Each node can determine its own cluster.
   b) Each node knows its on-hop neigbour. The clustering information (degree, ID, weight or status) of a node can be piggybacked in a periodic hello message.
   c) No cluster head is directly linked. Gateways and distributed gateways are the bridges between cluster heads.

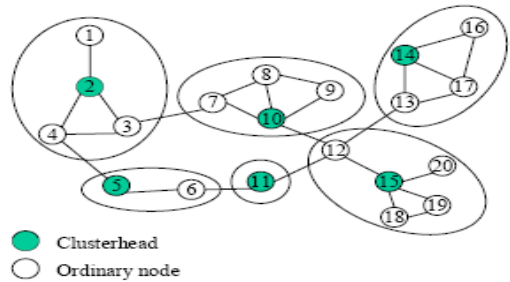

**Figure 3 :- System Model**

56

## 4. Checkpointing Algorithm

An ad hoc network does not have any predefined set up or structure. Nodes may always be moving and there may be frequent link failures. Each node acts as a router to pass the message. When a node fails, all other nodes learn the failure in finite time. We assume that the checkpointing algorithm operates both intra-cluster and inter-cluster. Nodes are referred to as process. Consider a cluster having a set of n nodes { N1, N2……N$_i$} involved in the execution of the algorithm. Each node $N_i$ maintains a dependency vector $Dv_i$ of size n which is initially empty and an entry $Dv_i[j]$ is set to 1 when $N_i$ receives since its last checkpoint at least one message from $N_j$. It is reset to 0 again when Node $N_i$ takes a checkpoint. Each node $N_i$ maintains a checkpoint sequence number $csn_i$. This $csn_i$ actually represents the current checkpointing interval of node $N_i$. The ith checkpoint interval of a process denotes all the computation performed between its $i^{th}$ and $(i+1)^{th}$ checkpoint, including the $i^{th}$ checkpoint but not the $(i+1)^{th}$ checkpoint. The $csn_i$ is initially set to 1 and is incremented when node $N_i$ takes a checkpoint. In this approach, we assume that only one node can initiate the chekpointing algorithm and that is the cluster head. This node is call as initiator node or cluster head. We define that process $N_k$ is dependent on another process $N_r$, if process $N_r$ since its last checkpoint has received at least one application message from process $N_k$. In our proposed scheme, we assume primary and secondary checkpoint request exchanges between cluster head and rest n-1 ordinary nodes. A permanent checkpoint request is denoted by $R_i(i=csni)$ where i is the current checkpoint sequence number of cluster head that initiates the checkpointing algorithm. It is sent by the initiator process $N_j$ to all its dependent nodes asking them to take their repective checkpoints. A tentative checkpoint request denoted by $R_{si}$ is sent from process $N_m$ to process $N_n$ which is dependent on $N_m$ to take a checkpoint $R_{si}$ means to its receiver process that iis the current checkpoint sequence number of the sender process. When $P_i$ sendsm to $P_j$, $P_i$ piggybacks c-state$_i$, own_csn$_i$ alongwith m. c_state$_i$ A flag. Set to '1' on the receipt of the minimum set. Set to '0' on receiving *commit* or *abort*. own_csn is the csn of $P_i$ at the time of sending m.

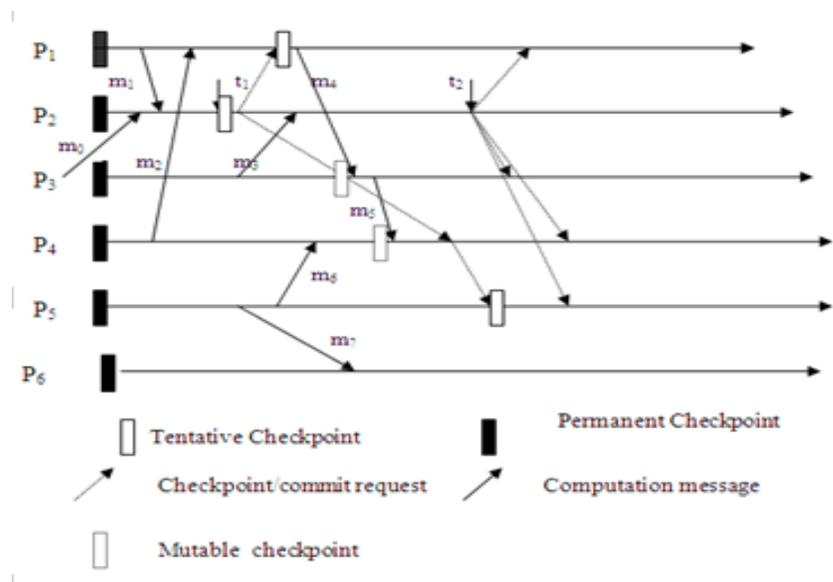

**Figure 4 :An example showing the Execution of the Proposed Protocol**

We explain our checkpointing algorithm with the help of an example. In Figure 4, at time $t_1$, $P_2$ initiates checkpointing process. $Dv_2[1]=1$ due to $m_1$; and $Dv_1[4]=1$ due to $m_2$. On the receipt of $m_0$, $P_2$ does not set $Dv_2[3] =1$, because, $P_3$ has taken permanent checkpoint after sending $m_0$. We assume that $P_1$ and $P_2$ are in the cell of the same Cluster, say $Cluster_{in}$. $Cluster_{in}$ computes minset (subset of minimum set) on the basis of Dv vectors maintained at $Cluster_{in}$, which in case of figure 3 is $\{P_1, P_2, P_4\}$. Therefore, $P_2$ sends checkpoint request to $P_1$ and $P_4$. After taking its tentative checkpoint, $P_1$ sends $m_4$ to $P_3$. $P_3$ takes mutable checkpoint before processing $m_4$. Similarly, $P_4$ takes mutable checkpoint before processing $m_5$. When $P_4$ receives the checkpoint request, it finds that it has already taken the mutable checkpoint; therefore, it converts its mutable checkpoint into tentative one. $P_4$ also finds that it was dependent upon $P_5$ before taking its mutable checkpoint and $P_5$ is not in the minimum set. Therefore, $P_4$ sends checkpoint request to $P_5$. At time $t_2$, $P_2$ receives responses from all relevant processes and sends the commit request along with the minimum set [$\{P_1, P_2, P_4, P_5\}$] to all processes. When a process, in the minimum set, receives the commit message, converts its tentative checkpoint into permanent one. When a process, not in the minimum set, receives the commit message, it discards its mutable checkpoint, if any. For the sake of simplicity, we have explained our algorithm with two-phase scheme.

When a process sends a computation message, it appends its own csn with it. When Pi receives m from Pj such that m. csn<=csn[j], the message is processed and no checkpoint is taken. Otherwise, it means that Pj has taken a checkpoint in the current initiation before sending m. Pi checks the following conditions:

1. $P_j$ was in the checkpointing state before sending m
2. $P_i$ has sent at least one message since last checkpoint
3. $P_i$ is not in checkpointing state while receiving m
4.

If all of these conditions are satisfied, Pi takes its induced checkpoint before processing *m*. If only conditions 1 and 3 are satisfied, P*i* updates its own *csn* before processing m. In other cases, message is processed. On the receipt of a message, D*v*[] of the receiver is updated.

### 4.1 Algorithm

We define the pseudo code here.

**Initiator Node N*i* (we call it as cluster head also)**

1. Take a checkpoint, check the dependency vector $DV_i[]$;
2. when $DV_i[k]= = 1$ for $1<=k<=n$
   send primary request – Rn to node Nk;
   /* checks dependency vector and sends checkpoint request */
3. increment the checkpoint sequence number $csn_i$;
4. continue normal computation;
   if any tentative checkpoint checkpoint request is received
   discard it and continue normal execution;
   Any node $N_j$ j!=I and $1<=j<=n$
   If $N_j$ receives a Permanent checkpoint request from $N_i$
   Take a checkpoint;
   /* if $N_j$ is busy with other high priority job, it takes checkpoint after completing the job; otherwise takes checkpoint immediately */
   If $DV_j[\ ]$=null;
   Increment $csn_j$;
   *Continue computation;*
   Else
   send secondary checkpoint request to each of $N_k$ such that $DV_j[k]=1$;
   increment $csn_j$;

```
continue computation;
else if N_j receives a secondary checkpoint request
ifN_j has already participated in the checkpoint algorithm
ignore the checkpoint request and continue computation;
else
take a checkpoint;
/* if N_j is busy with other high priority job, it takes checkpoint after completing the job;*/
ifDV_j[]=null;

incrementcsn_j;
continue computation;
else
sendtentative checkpoint request to each N_k such that DV_j[k]=1;
incrementcsn_j;
continue computation;
else if N_j receives piggybacked application message
If N_j has already participated in the checkpointing algorithm
/* csn_j is greater than the received checkpoint sequence number */
process the message and continue computation
else
/* if N_j is busy with other high priority job, it takes checkpoint after completing the job;
otherwise takes checkpoint immediately */
ifDV_j[]=null;
incrementcsnj;
process the message;
continue computation;
else
sendtentative checkpoint request to each N_k such that DV_j[k]=1;
incrementcsn_j;
process the message;
continue computation;
```

Consider the pseudo code for any node $N_j$. Node $N_j$ makes sure that all processes from which it has received messages also take checkpoints so that there are no orphan messages that it has received. Also, the node $N_j$ first takes its checkpoint if needed, then processes the received piggybacked application message. Thus, such messages cannot be an orphan. Hence, algorithm generates a consistent global state.

### 4.2 Proof of correctness

Let $GC_i = \{C_{1,x}, C_{2,y}, ............, C_{n,z}\}$ be some consistent global state created by our algorithm, where $C_{i,x}$ is the $x^{th}$ checkpoint of $P_i$.

### Theorem I: The global state created by the $i^{th}$ iteration of the checkpointing protocol is consistent.

**Proof:** Let us consider that the system is in consistent state when a process initiates checkpointing. The recorded global state will be inconsistent only if there exists a message $m$ between two processes $P_i$ and $P_j$ such that $P_i$ sends $m$ after taking the checkpoint $C_{i,x}$, $P_j$ receives $m$ before taking the checkpoint $C_{j,y}$, and both $C_{i,x}$ and $C_{j,y}$ are the members of the new global state. We prove the result by contradiction that no such message exists. We consider all four possibilities as follows:

**Case I:** *$P_i$ belongs to the minimum set and $P_j$ does not:*

As $P_i$ is in the minimum set, $C_{i,x}$ is the checkpoint taken by $P_i$ during the current initiation and $C_{j,y}$ is the checkpoint taken by $P_j$ during some previous initiation i.e. $C_{j,y}$ $C_{i,x}$. Therefore rec(m) $C_{j,y}$ and $C_{i,x}$ send(m) implies rec(m) $C_{j,y}$ $C_{i,x}$ send(m) implies rec(m) send(m) which is not possible. ' ' is the Lamport's happened before relation.

**Case II:** *Both $P_i$ and $P_j$ are in minimum set:*

Both $C_{i,x}$ and $C_{j,y}$ are the checkpoints taken during current initiation. There are following possibilities:
*(a) $P_i$ sends m after taking its mutable or tentative checkpoint and $P_j$ receives m before taking its tentative/mutable checkpoint:*
When $P_i$ takes its mutable/tentative checkpoint, it increments own_csn$_i$, sets c_state$_i$. As it sends m after taking the checkpoint, it will piggyback updated own_csn$_i$ and c_state$_i$ with m. If $P_j$ receives m before taking its mutable/tentative checkpoint, the following condition will be true at the time of receiving m:

$$((m.own\_csn > csn[i]) \wedge (c\_state_j == 0) \wedge (m.c\_state == 1) \wedge (send_j))$$

In this case, $P_j$ will take its mutable checkpoint before receiving m. It should be noted that (send$_j$) will also be true at the time of receiving m. Otherwise, it means that $P_j$ has not sent any message since last permanent checkpoint. In this case, $P_j$ will process m without taking any checkpoint and it will not be included in the minimum set in any case.

*(b) $P_i$ sends m after commit and $P_j$ receives m before taking tentative checkpoint:*
As $P_j$ is in minimum set, initiator can issue a commit only after $P_j$ takes tentative checkpoint and informs initiator. Therefore the event rec(m) at $P_j$ cannot take place before $P_j$ takes the checkpoint.

**Case III:** *$P_i$ is not in minimum set but $P_j$ is in minimum set:*

Checkpoint $C_{j,y}$ belongs to the current initiation and $C_{i,x}$ is from some previous initiation. In this case, when $P_j$ takes the checkpoint, it will ensure that checkpoint request has been sent to $P_i$. When $P_j$ takes its tentative checkpoint, $Dv_j[i]=1$ due to receive of m. In this case, if $P_i$ is not in the computed minimum set so far, $P_j$ will send the checkpoint request to $P_i$.

**Case IV:** *Both $P_i$ and $P_j$ are not in minimum set:*

Neither $P_i$ nor $P_j$ will take a new checkpoint, therefore, no such *m* is possible unless and until it already exists.

**Theorem II:** A process can not be a member of any minimum set, if it has not sent a message inits current checkpointing interval and minimum numbers of processes take checkpoint.
**Proof:** if $N_i$ is initiator and initiates its checkpointing algorithms then it contains all theprocesses in *minset* which are directly or transitively dependent on $N_i$ in current checkpointinitiation. So a process can not become the member of *minset* if it has not sent a message incurrent checkpoint initiation. Process $N_j$ will take a checkpoint related to current initiation if andonly if it is directly or indirectly dependent or sends a computation message to a process whichis directly or indirectly dependent on initiator. If dependency is counted at the time of initiationit will get the checkpoint request directly from the initiator and in case of any tardy message itwill get the checkpoint request from that particular process in which it is directly dependent.On the basis of theorem II we conclude that:

a) Each process notified by the global initiator or any process which are directly orindirectly after taking the checkpoint at most one checkpoint.
b) Our algorithms only forces minimal number of processes to take checkpoint.
c) If the set of checkpoint the checkpoint state is consistent before execution of ourproposed algorithm, then it also consistent after the termination of algorithm.

## 5. Performance comparison

We compare our work with [7], [22], [9] and [10]. In [7], a cluster takes two types ofcheckpoints; processes inside a cluster take checkpoints synchronously and a cluster takes a communication induced checkpoint whenever it receives an intercluster application message. Each cluster maintains a sequence number (SN). SN is incremented each time a cluster level checkpoint is committed. Each cluster also maintains a Dv (Direct dependency vector) with a size equal to the number of clusters in the cluster federation. Whenever a cluster (i.e. a process in it) fails, after recovery it broadcasts an alert message with the SN of the failed cluster. This alert message triggers the next iteration of the algorithm. All other clusters, on receiving this alert message decide if they need to roll back by checking the corresponding entries in the Dv vectors. This algorithm has the following advantage; simultaneous execution of the algorithm by all participating clusters contributes to its speed of execution. However, the main drawback of the algorithm is that if we consider a particular message pattern where all the clusters have to roll back except the failed cluster, then all the clusters have to send alert messages to every other cluster. This results in a message storm. But in our approach when a process of a cluster fails it broadcasts just one control message for link failure.

In [22], the authors have addressed the need of integrating independent and coordinated checkpointing schemes for applications running in a hybrid distributed environment containing multiple heterogeneous subsystems. This algorithm mainly works as follows – Firstly, it states that, independent checkpoint subsystem takes a new coordinated checkpoint set if it sends an intercluster application message. Secondly, it states that, a process Pi of independent checkpointing subsystem takes a new independent checkpoint before processing an already received intercluster application message, if Pi has sent any intracluster application message after taking its last checkpoint. So, if the independent checkpointing subsystem has sent k number of intercluster application messages in a time period T, then it has to take k number of coordinated checkpoint sets besides the regular local checkpoints taken asynchronously by its processes. In our approach, if we consider the same situation, only the minimum number of processes takes checkpoints. So we reduce drastically the number of checkpoints to be taken by the cluster subsystem.

In [9] Cao-Singhal proposed a mutable checkpoint based non-blocking minimum-processcoordinated checkpointing algorithm. This algorithm completes its processing in the followingthree steps. First initiator MSS sends tentative checkpoint request to minimum number ofprocesses that need to take checkpoint. Secondly MSS gets the acknowledgement from all processes to whom it sent checkpointrequest. At last MSSsends the commit request to convert its tentative checkpoint into permanent. Thus algorithm is non-blocking and minimum process but suffer fromuseless checkpoints.

In [10], P.Kumar et al. also proposed minimum process coordinated checkpoint algorithm for mobile system. The algorithm suffers from uselesscheckpoint.

Our proposed approach is quite different from all the above mentioned approaches. Firstly, our approach considers a cluster based protocols from the class of ad hoc networks whereas in the above mentioned three approaches general concept of mobile computing is considered.Seondly, our approach also explains the recovery process of the ordinary nodes and cluster head. Lastly, our proposed algorithm generates the consistent global state without using any useless checkpoint, it is non-blocking and it is applied on the ad hoc networks.

**Table 1 : - Comparison with the related work**

| Parameters | Compariosn with [7] | Comparison with [22] | Comparison with [9] | Comparison with [10] | Our algorithm |
|---|---|---|---|---|---|
| **Non-blocking** | Yes | Yes | Yes | Yes | Yes |
| **Minimum Process** | No | No | Yes | Yes | Yes |
| **Supports MANET's** | Yes | Yes | No | No | Yes |
| **Number of checkpoints** | Less | More | Less | Less | Less |
| **No. of control messages** | More | More | Less | Less | Less |

## 6. Conclusion

When designing an efficient ad hoc network application, we must consider the resource constraints and their scalability. Ad hoc network users concerned about information quality and user requirements for real-time features are also increasing. Moreover, ad hoc network applications are expanding into harsher and more dangerous environments. Therefore, checkpointing schemes have emerged as an important issues.

Clustering routing protocols such as CBRP are designed to improve both energy efficiency and scalability. These protocols compose clusters and elect a cluster head in each cluster. The cluster heads aggregate data from its member nodes and reduces the amount of messages sent by member nodes to the BS directly. In clustering routing protocol, cluster head management is needed because the role of cluster head is more important than other member nodes.

In this paper, we have proposed a minimum process and non-blocking checkpointing scheme for clustering routing protocols. The main features of our algorithm are 1) it caters the needs of ad hoc environment ; 2) minimum number of processes take the checkpoint. Also, our scheme minimizes the number of control messages needed and also take no useless checkpoints. And finally, it reduces the energy consumption and recovery latency when a cluster head fails.

## REFRENCES